\documentclass[conference]{IEEEtran}
\IEEEoverridecommandlockouts
\usepackage{cite}
\usepackage{amsmath,amssymb,amsfonts,graphicx,tabularx}
\usepackage{algorithmic}
\usepackage{graphicx}
\usepackage{textcomp}
\usepackage{xcolor}
\usepackage{float}
\usepackage{comment}
\graphicspath{ {./images/} }
\def\BibTeX{{\rm B\kern-.05em{\sc i\kern-.025em b}\kern-.08em
    T\kern-.1667em\lower.7ex\hbox{E}\kern-.125emX}}
\begin{document}

\title{Extracting Task Trees Using Knowledge Retrieval Search Algorithms in Functional Object-Oriented Network\\
}

\author{\IEEEauthorblockN{1\textsuperscript{st} Tyree Lewis}
\IEEEauthorblockA{\textit{Department of Computer Science and Engineering} \\
\textit{University of South Florida}\\
Tampa FL, United States \\
tlewis10@usf.edu}
}

\maketitle

\begin{abstract}
The functional object-oriented network (FOON) has been developed as a knowledge representation method that can be used by robots in order to perform task planning. A FOON can be observed as a graph that can provide an ordered plan for robots to retrieve a task tree, through the knowledge retrieval process. We compare two search algorithms to evaluate their performance in extracting task trees: iterative deepening search (IDS) and greedy best-first search (GBFS) with two different heuristic functions. Then, we determine which algorithm is capable of obtaining a task tree for various cooking recipes using the least number of functional units. Preliminary results show that each algorithm can perform better than the other, depending on the recipe provided to the search algorithm.
\end{abstract}


\section{Introduction}
Over the years, research has been conducted to find different approaches to help robots learn about their environment and the objects that exist in that space. A knowledge representation called the function object-oriented network (FOON) was one such development to model the ways humans observe and perform tasks into a framework that can be applied to the field of robotics. To advance the capabilities of intelligent and autonomous robots, this representation aims to bridge the gap between understanding tasks and action between humans and robots \cite{b1}. Through the understanding about human interaction with objects in their daily lives, studies have found the motor response of performing an action has a relationship between the manipulated object and the interacted object \cite{b2}. The problem, however, is that while humans are able to adapt to scenarios where there is limited knowledge about available objects in a system to perform actions, a robot requires fully detailed instructions to perform certain tasks with provided objects, and would not be able to execute actions if objects are unavailable \cite{b3}. As a result, developing search algorithms for FOON is critical to derive knowledge about necessary objects in a scenario from various existing subgraphs combined into task trees, so that a robot will always have the necessary information to perform actions. Furthermore, it allows for robots to be adapted to various systems, by taking object and state information from existing task trees and use them in other task trees where that knowledge may be unavailable.
\newline \indent
In the work done by \cite{b3}, they tackle the challenge of flexible task planning with robots. It was observed that when robots are faced with unknown or limited knowledge, they are unable to creatively adapt their task plans in order to complete an action. To resolve this issue, knowledge from 140 cooking recipes were structured into FOON knowledge graphs. Providing this information to robots allows them learn, by making semantic similarities between the objects and states needed to complete actions. Their results showed their system was able to provide task sequences with an accuracy of 76\%. Evaluations made by \cite{b4}, showed that even when providing robots with the necessary FOON knowledge graphs, there may still be tasks that are more optimal for humans to perform compared to robots. Complex manipulation tasks, such as performing cooking with a provided recipe, may involve several steps that are risky for a robot to implement. This presents a challenge from reaching the state where robots can be fully autonomous with their actions. As such, this work proposed adding weights to FOONs in order to allocate manipulation action load between a robot and human to improve performance, while minimizing human effort. From the results, it showed that there are several examples where weighted FOON and task planning algorithm are able to distribute work load between a robot and human with higher success rates to complete the tasks. Due to the importance of knowledge retrieval through graph search to obtain the information necessary for robots to learn, \cite{b5} identifies the issue of limited video sources to acquire data to be generated into a FOON. Two means were developed to generalize knowledge to be applied to similar objects in a FOON without requiring manually annotating new sources for information. They discuss using object similarity to create new functional units and compressing functional units by object categories instead of particular objects.

\section{FOON creation and Video annotation}
\subsection{FOON creation}
Before discussing the process of creating a FOON using video annotation, there are several terminologies that are based around the concept of a FOON:

\begin{itemize}
  \item \textbf{Functional unit:} The combination of object and motion nodes to reflect a single action 
  \item \textbf{Subgraph:} A FOON based on a single activity
  \item \textbf{Universal FOON:} A FOON created from merging two or more subgraphs containing several sources of information
  \item \textbf{Task tree:} The resulting subgraph from knowledge retrieval
  \item \textbf{Object node:} The objects and states before and after the motion node in a functional unit
  \item \textbf{Motion node:} The single action in a functional unit that changes the state(s) of input object nodes
\end{itemize}

\begin{figure}[H]
\centering
\includegraphics[scale=0.55]{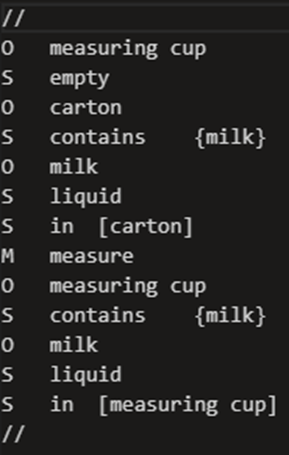}
\caption{Example Functional Unit.}
\label{fig:fu_sample}
\end{figure}

\begin{figure}[H]
\centering
\includegraphics[scale=0.55]{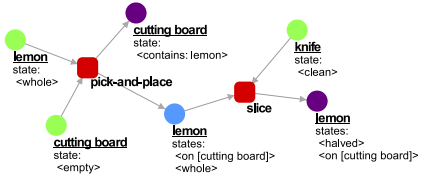}
\caption{Two Connected Functional Units \cite{b6}.}
\label{fig:two_fus}
\end{figure}

\begin{figure}[H]
\centering
\includegraphics[scale=0.20]{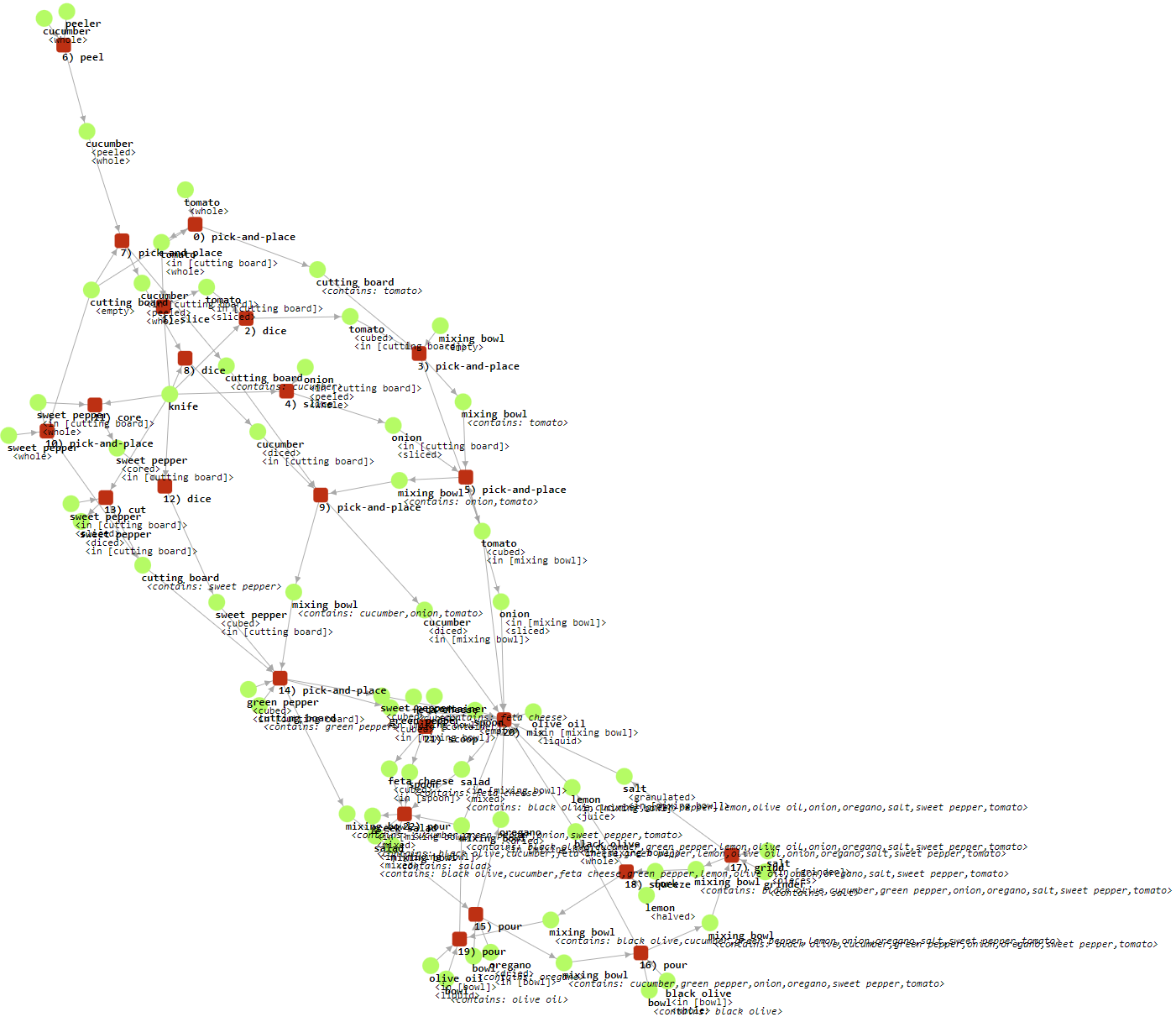}
\caption{Subgraph of a Greek Salad Recipe.}
\label{fig:greek_salad_subgraph}
\end{figure}

\begin{figure}[H]
\centering
\includegraphics[scale=0.24]{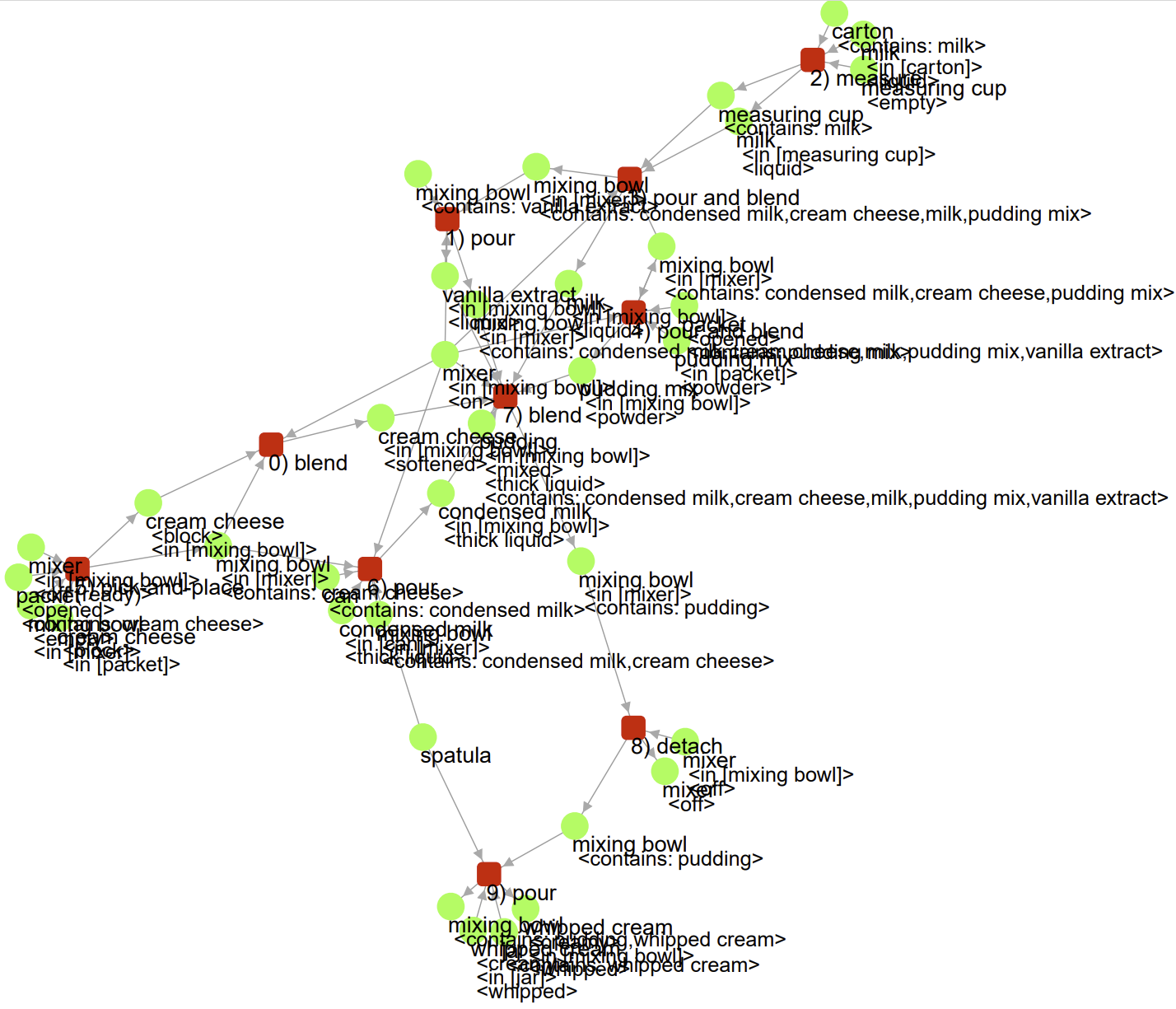}
\caption{Task Tree of a Greek Salad Recipe Using IDS Search Algorithm.}
\label{fig:greek_salad_task_tree}
\end{figure}

\begin{figure}[H]
\centering
\includegraphics[scale=0.5]{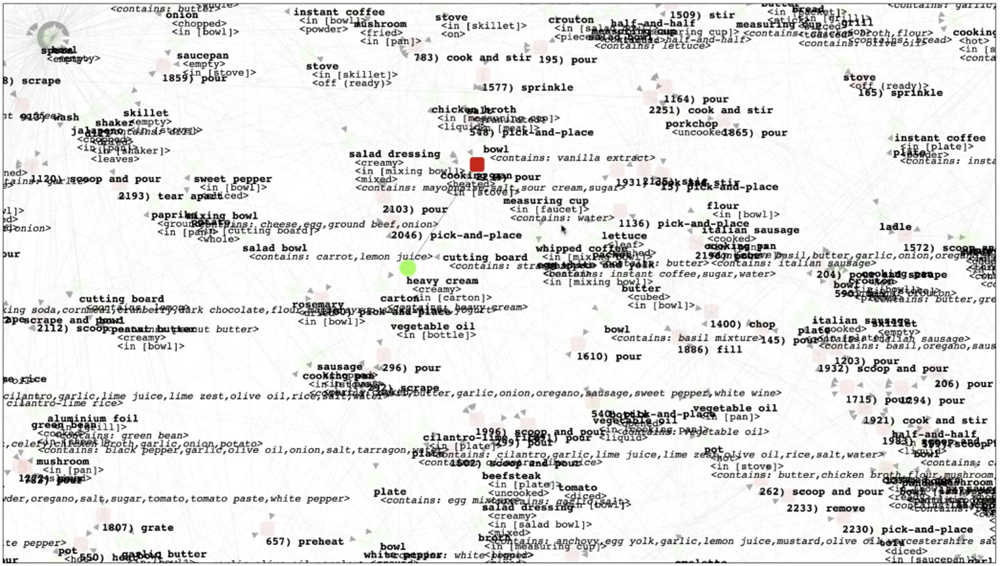}
\caption{A Universal Foon.}
\label{fig:universal_foon}
\end{figure}

Figure \ref{fig:fu_sample} is an example of a functional unit. In this case, there are three input object nodes that appear before the motion node, M. After the motion node, there are two output object nodes. Another representation can be seen in Figure \ref{fig:two_fus}, where in this example there are two connected functional units. The output nodes are shaped as circles, and the green nodes are input object nodes and the purple nodes are output object nodes. If an object node is both an input and output node, it is shown in the Figure in blue. Lastly, the motion nodes appear as red squares. In total, there are three input object nodes, two output object nodes, one object node that is both an input and output object node, and two motion nodes in this example. \newline \indent A completed subgraph is shown in Figure \ref{fig:greek_salad_subgraph}, which is an example of preparing a greek salad dish. Within the subgraph are multiple functional units based on this single recipe. Figure \ref{fig:greek_salad_task_tree} shows an extracted task tree created from an Iterative Deepening Search algorithm when passing in the goal node from the greek salad subgraph. From the two figures, it can be observed that the subgraph of the greek salad recipe is larger in size and different in appearance than the extracted task tree, because the search uses a specific algorithm to get all the nodes needed to create it. When this greek salad subgraph is merged with all the other cooking recipes, a universal FOON is formed, shown in Figure \ref{fig:universal_foon}.

\subsection{Video Annotation}
\begin{figure}[H]
\centering
\includegraphics[scale=0.65]{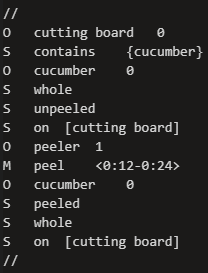}
\caption{Functional Unit With Timestamp in Motion Node.}
\label{fig:fu_time}
\end{figure}

The process of creating a FOON involves retrieving existing knowledge and representing that information in a structure, based on object-motion affordances \cite{b5}. A FOON is manually generated by hand using annotations from video demonstrations, such as cooking videos from YouTube, and they are converted into the FOON graph structure \cite{b6}. The annotation process begins, by recording the actions, objects, and state changes in the form of functional units. The motion node has a timestamp recorded when a particular motion took place to output the resulting output object node, as shown in Figure \ref{fig:fu_time}. Creating a series of functional units throughout the start and end of a cooking recipe video creates a subgraph for that recipe. This completed subgraph can be merged into a Universal FOON to be used for task tree retrieval.

\section{Methodology}
Two search algorithms were implemented to extract a task tree to prepare dishes that exist in FOON. The general process requires starting from the goal node and search for candidate functional units. For each candidate unit, a search is performed on its input nodes to determine if they exist in the kitchen. If the node does not exist in the kitchen, it is explored. The search concludes when all nodes are already in the kitchen. Based on the algorithm, different candidate units can be selected and the resulting task tree will vary per dish.
\raggedbottom
\subsection{Iterative Deepening Search}
\begin{figure}[H]
\centering
\includegraphics[scale=0.52]{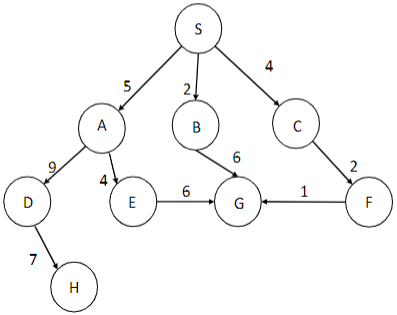}
\caption{Example Graph for IDS Provided by the University of South Florida.}
\label{fig:graph_IDS}
\end{figure}

\begin{figure}[H]
\centering
\includegraphics[scale=0.52]{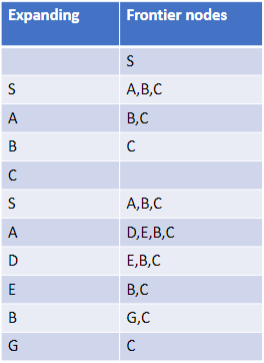}
\caption{Solution to Example Graph for IDS Provided by the University of South Florida.}
\label{fig:solution_IDS}
\end{figure}

The first search implemented is called Iterative Deepening Search (IDS). It is a solution to the drawbacks of Depth First Search (DFS), by first defining a depth bound, then incrementally changing its depth bound until the goal is found. At each depth bound, DFS is performed. An example is shown based on the graph in Figure \ref{fig:graph_IDS} and a solution using IDS in Figure \ref{fig:solution_IDS}. In this example, all children at the start of the node are treated as leaves. If the goal is not found at the current depth, DFS is restarted and the depth bound is increased by 1.

Similarly to this example, in the implementation used for the dishes that exist in FOON, a depth bound was initialized to 0. Starting from the goal node, the process of searching for input nodes for a candidate unit is done only at the current depth bound. If the goal is not found, meaning there are still items to search, the max depth bound is increased by 1, and DFS is repeated starting from the first depth until the current max depth.  

\subsection{Greedy Best-First Search}
Greedy Best-First Search (GBFS) is performed by sorting nodes by increasing values of an evaluation function, f(n) = h(n). This means choosing a candidate unit, based on heuristic functions for the dishes. Heuristic functions are defined as h(n), which is method used to guide the search algorithm. They are used to represent how close n is to the goal and is also the minimum cost path from node n to the goal. Depending on the heuristic function chosen, different candidate units may be selected, which changes the resulting task trees. For the dish preparation, the two heuristic functions implemented are as follows:

\begin{enumerate}
	    \item \textbf{h(n) = success rate of the motion}
	    \item \textbf{h(n) = number of input objects in the functional unit}
\end{enumerate}

For the first heuristic function, there were a list of motions with success rates provided in a motion.txt file. When determining the path, there are several potential candidate units to select. Each candidate unit has a motion node, which corresponds to a certain success rate. The candidate unit chosen was the one that gave the highest success rate for executing that particular motion successfully. The second heuristic function required evaluating the number of input objects per potential candidate unit. To count the number of input objects, both this number and the number of ingredients used for that input object were totalled. The path with the least number of input objects used was the candidate unit selected.

\section{Experiment/Discussion}
To evaluate the performance of the different search algorithms, five goal nodes were used to start the search to extract task trees, which were created in a file named goal\_nodes.json. A python script named search.py was used to perform the task tree retrieval. Three different functions named after each of the search algorithms were implemented in this file to get the experimental results. The following goal nodes were searched for in this list below:

\begin{itemize}
	    \item \textbf{Goal \#1:} Whipped Cream
	    \item \textbf{Goal \#2:} Greek Salad
	    \item \textbf{Goal \#3:} Macaroni
	    \item \textbf{Goal \#4:} Sweet Potato
	    \item \textbf{Goal \#5:} Ice
	    \item \textbf{Goal \#6:} Fried Rice
	    \item \textbf{Goal \#7:} Enchilada
\end{itemize}

Based on the results in Table \ref{tab:ice}, IDS was the better search algorithm, as it took ten functional units in order to reach the goal. For Table \ref{tab:greek_salad}, IDS has the best performance when searching for the greek salad goal, but the other two algorithms had similar numbers of functional units. Both IDS and GBFS with heuristic function \#2 had the same performance in Table \ref{tab:macaroni}. Tables \ref{tab:sweet_potato} and \ref{tab:ice} shows all three algorithms extracted search trees had the same number of functional units. Compared to the previous tables, Table \ref{tab:fried rice} shows that GBFS with heuristic \#2 was the better search algorithm compared to the others. Lastly, GBFS with heuristic \#1 was significantly better in Table \ref{tab:enchilada} compared to the other search algorithms.

\begin{table}[H]
\centering
\begin{tabular}{|cc|}
\hline
\multicolumn{2}{|c|}{\textbf{Goal \#1: Whipped Cream}}                                                          \\ \hline
\multicolumn{1}{|c|}{\textbf{Search Algorithm}}      & \multicolumn{1}{l|}{\textbf{Number of Functional Units}} \\ \hline
\multicolumn{1}{|c|}{\textbf{IDS}}                   & 10                                                       \\ \hline
\multicolumn{1}{|c|}{\textbf{GBFS with Heuristic 1}} & 15                                                       \\ \hline
\multicolumn{1}{|c|}{\textbf{GBFS with Heuristic 2}} & 15                                                       \\ \hline
\end{tabular}
\caption{Number of Functional Units for Creating Task Trees for Goal \#1 By Search Algorithm.}
\label{tab:whipped_cream}
\end{table}

\begin{table}[H]
\centering
\begin{tabular}{|cc|}
\hline
\multicolumn{2}{|c|}{\textbf{Goal \#2: Greek Salad}}                                                          \\ \hline
\multicolumn{1}{|c|}{\textbf{Search Algorithm}}      & \multicolumn{1}{l|}{\textbf{Number of Functional Units}} \\ \hline
\multicolumn{1}{|c|}{\textbf{IDS}}                   & 28                                                       \\ \hline
\multicolumn{1}{|c|}{\textbf{GBFS with Heuristic 1}} & 33                                                       \\ \hline
\multicolumn{1}{|c|}{\textbf{GBFS with Heuristic 2}} & 31                                                       \\ \hline
\end{tabular}
\caption{Number of Functional Units for Creating Task Trees for Goal \#2 By Search Algorithm.}
\label{tab:greek_salad}
\end{table}

\begin{table}[H]
\centering
\begin{tabular}{|cc|}
\hline
\multicolumn{2}{|c|}{\textbf{Goal \#3: Macaroni}}                                                          \\ \hline
\multicolumn{1}{|c|}{\textbf{Search Algorithm}}      & \multicolumn{1}{l|}{\textbf{Number of Functional Units}} \\ \hline
\multicolumn{1}{|c|}{\textbf{IDS}}                   & 7                                                       \\ \hline
\multicolumn{1}{|c|}{\textbf{GBFS with Heuristic 1}} & 8                                                       \\ \hline
\multicolumn{1}{|c|}{\textbf{GBFS with Heuristic 2}} & 7                                                       \\ \hline
\end{tabular}
\caption{Number of Functional Units for Creating Task Trees for Goal \#3 By Search Algorithm.}
\label{tab:macaroni}
\end{table}

\begin{table}[H]
\centering
\begin{tabular}{|cc|}
\hline
\multicolumn{2}{|c|}{\textbf{Goal \#4: Sweet Potato}}                                                          \\ \hline
\multicolumn{1}{|c|}{\textbf{Search Algorithm}}      & \multicolumn{1}{l|}{\textbf{Number of Functional Units}} \\ \hline
\multicolumn{1}{|c|}{\textbf{IDS}}                   & 3                                                       \\ \hline
\multicolumn{1}{|c|}{\textbf{GBFS with Heuristic 1}} & 3                                                       \\ \hline
\multicolumn{1}{|c|}{\textbf{GBFS with Heuristic 2}} & 3                                                       \\ \hline
\end{tabular}
\caption{Number of Functional Units for Creating Task Trees for Goal \#4 By Search Algorithm.}
\label{tab:sweet_potato}
\end{table}

\begin{table}[H]
\centering
\begin{tabular}{|cc|}
\hline
\multicolumn{2}{|c|}{\textbf{Goal \#5: Ice}}                                                          \\ \hline
\multicolumn{1}{|c|}{\textbf{Search Algorithm}}      & \multicolumn{1}{l|}{\textbf{Number of Functional Units}} \\ \hline
\multicolumn{1}{|c|}{\textbf{IDS}}                   & 1                                                       \\ \hline
\multicolumn{1}{|c|}{\textbf{GBFS with Heuristic 1}} & 1                                                       \\ \hline
\multicolumn{1}{|c|}{\textbf{GBFS with Heuristic 2}} & 1                                                       \\ \hline
\end{tabular}
\caption{Number of Functional Units for Creating Task Trees for Goal \#5 By Search Algorithm.}
\label{tab:ice}
\end{table}

\begin{table}[H]
\centering
\begin{tabular}{|cc|}
\hline
\multicolumn{2}{|c|}{\textbf{Goal \#6: Fried Rice}}                                                          \\ \hline
\multicolumn{1}{|c|}{\textbf{Search Algorithm}}      & \multicolumn{1}{l|}{\textbf{Number of Functional Units}} \\ \hline
\multicolumn{1}{|c|}{\textbf{IDS}}                   & 38                                                       \\ \hline
\multicolumn{1}{|c|}{\textbf{GBFS with Heuristic 1}} & 35                                                       \\ \hline
\multicolumn{1}{|c|}{\textbf{GBFS with Heuristic 2}} & 33                                                       \\ \hline
\end{tabular}
\caption{Number of Functional Units for Creating Task Trees for Goal \#6 By Search Algorithm.}
\label{tab:fried rice}
\end{table}

\begin{table}[H]
\centering
\begin{tabular}{|cc|}
\hline
\multicolumn{2}{|c|}{\textbf{Goal \#7: Enchilada}}                                                          \\ \hline
\multicolumn{1}{|c|}{\textbf{Search Algorithm}}      & \multicolumn{1}{l|}{\textbf{Number of Functional Units}} \\ \hline
\multicolumn{1}{|c|}{\textbf{IDS}}                   & 43                                                       \\ \hline
\multicolumn{1}{|c|}{\textbf{GBFS with Heuristic 1}} & 15                                                       \\ \hline
\multicolumn{1}{|c|}{\textbf{GBFS with Heuristic 2}} & 39                                                       \\ \hline
\end{tabular}
\caption{Number of Functional Units for Creating Task Trees for Goal \#7 By Search Algorithm.}
\label{tab:enchilada}
\end{table}

Table \ref{tab:complexity} shows the computer and memory complexity of each search algorithm to extract the task trees, where b is the branching factor, m is the maximum depth of the search tree, and d is the depth. 

\begin{table}[H]
\centering
\begin{tabular}{|c|c|c|}
\hline
\textbf{Search Algorithm}      & \multicolumn{1}{l|}{\textbf{Computer Complexity}} & \multicolumn{1}{l|}{\textbf{Memory Complexity}} \\ \hline
\textbf{IDS}                   & O(b\textsuperscript{d})                           & O(d)                                            \\ \hline
\textbf{GBFS with Heuristic 1} & O(bm)                                             & O(d)                                            \\ \hline
\textbf{GBFS with Heuristic 2} & O(bm)                                             & O(d)                                            \\ \hline
\end{tabular}
\caption{Computer and Memory Complexity of Each Search Algorithm}
\label{tab:complexity}
\end{table}

The resulting computer and memory complexity for GBFS, is because this is not an optimal search. The path chosen based on the selected candidate unit may not be the optimal path. For IDS, it depends on how far in the depth is required to find the goal.

\vspace{12pt}


\begin{thebibliography}{00}
\bibitem{b1} Paulius, D., \& Sun, Y. (2019). A survey of knowledge representation in service robotics. Robotics and Autonomous Systems, 118, 13-30.
\bibitem{b2} Paulius, D., Huang, Y., Milton, R., Buchanan, W. D., Sam, J., \& Sun, Y. (2016, October). Functional object-oriented network for manipulation learning. In 2016 IEEE/RSJ International Conference on Intelligent Robots and Systems (IROS) (pp. 2655-2662). IEEE.
\bibitem{b3} Sakib, M. S., Paulius, D., \& Sun, Y. (2022). Approximate Task Tree Retrieval in a Knowledge Network for Robotic Cooking. IEEE Robotics and Automation Letters, 7(4), 11492-11499.
\bibitem{b4} Paulius, D., Dong, K. S. P., \& Sun, Y. (2021, May). Task Planning with a Weighted Functional Object-Oriented Network. In 2021 IEEE International Conference on Robotics and Automation (ICRA) (pp. 3904-3910). IEEE.
\bibitem{b5} Paulius, D., Jelodar, A. B., \& Sun, Y. (2018, May). Functional object-oriented network: Construction \& expansion. In 2018 IEEE International Conference on Robotics and Automation (ICRA) (pp. 5935-5941). IEEE.
\bibitem{b6} Sakib, M. S., Baez, H., Paulius, D., \& Sun, Y. (2021). Evaluating recipes generated from functional object-oriented network. arXiv preprint arXiv:2106.00728.
\end{thebibliography}
\end{document}